\documentclass[3p,times,procedia]{elsarticle}
\usepackage{nupha_ecrc}
\usepackage{subcaption}
\usepackage{amsmath}
\usepackage{lineno}
\usepackage{floatrow}
\usepackage{mwe}
\usepackage{lipsum}
\volume{00}

\firstpage{1}

\journalname{Nuclear Physics A}

\runauth{Anthony Hodges for the PHENIX Collaboration}

\jid{nupha}

\jnltitlelogo{Nuclear Physics A}

\usepackage{amssymb}
\usepackage[figuresright]{rotating}

\begin{document}
\sloppy
\begin{frontmatter}

%% Title, authors and addresses

%% use the tnoteref command within \title for footnotes;
%% use the tnotetext command for the associated footnote;
%% use the fnref command within \author or \address for footnotes;
%% use the fntext command for the associated footnote;
%% use the corref command within \author for corresponding author footnotes;
%% use the cortext command for the associated footnote;
%% use the ead command for the email address,
%% and the form \ead[url] for the home page:
%%
%% \title{Title\tnoteref{label1}}
%% \tnotetext[label1]{}
%% \author{Name\corref{cor1}\fnref{label2}}
%% \ead{email address}
%% \ead[url]{home page}
%% \fntext[label2]{}
%% \cortext[cor1]{}
%% \address{Address\fnref{label3}}
%% \fntext[label3]{}

%% Instructions from Editor: Please use the following \dochead only in the preprint version (e-print arXiv etc.); 
%% use empty \dochead{} when submitting to Nuclear Physics A!
\dochead{XXVIIIth International Conference on Ultrarelativistic Nucleus-Nucleus Collisions\\ (Quark Matter 2019)}
%\dochead{}
%% Use \dochead if there is an article header, e.g. \dochead{Short communication}
%% \dochead can also be used to include a conference title, if directed by the editors
%% e.g. \dochead{17th International Conference on Dynamical Processes in Excited States of Solids}

\title{In-Medium Jet Modification Measured by PHENIX Via Two-Particle Correlations and High $p_T$ Hadrons in $A+A$ Collisions}

%% use optional labels to link authors explicitly to addresses:
%% \author[label1,label2]{<author name>}
%% \address[label1]{<address>}
%% \address[label2]{<address>}

\author{Anthony Hodges for the PHENIX Collaboration}

\address{Georgia State University}

\begin{abstract}
%% Text of abstract
The first evidence of jet quenching was observed at RHIC via suppression of single high $p_T$ hadron $R_{AA}$ and the disappearance of the away-side jet peak in two-particle correlations. Since then, hadron $R_{AA}$ and two-particle correlations continue to be useful probes of the QGP in heavy ion collisions, since the particles involved are fragments of the jets produced in the initial hard scattering. PHENIX recently improved the width measurements extracted from $\pi^0$-hadron correlations
after removing the higher order flow terms in the underlying event subtraction. Measurements of the away-side jet correlated with high $p_T$ neutral pions show an increase in low $p_T$ particle production at wide angles consistent with theoretical expectations for energy loss. The system size dependence of energy loss is further investigated at PHENIX by measuring the absolute yield and $R_{AA}$ for various hadron species at high $p_T$ in several collision systems including $U+U$ and $Cu+Au$. These proceedings will present the newest PHENIX $R_{AA}$ and two-particle correlation measurements and their role in our understanding of jet quenching and medium response in heavy ion collisions.
\end{abstract}

\begin{keyword}
%% keywords here, in the form: keyword \sep keyword
Jets \sep Two-Particle Correlations \sep RAA \sep Hadrons \sep Heavy Ion Collisions \sep Quark Gluon Plasma
%% MSC codes here, in the form: \MSC code \sep code
%% or \MSC[2008] code \sep code (2000 is the default)

\end{keyword}

\end{frontmatter}

%%
%% Start line numbering here if you want
%%
% \linenumbers

%% main text
\section{Introduction}
\label{Introduction}
High energy heavy ion collisions produced at the Relativistic Heavy Ion Collider (RHIC) and the Large Hadron Collider (LHC) provide a unique environment to study Quantum Chromodynamics (QCD). In heavy ion collisions, a hot, dense state of matter, known as a Quark Gluon Plasma (QGP), is formed in which quarks and gluons have their own degrees of freedom. After the initial hard scatterings in the early stages of the collision, the now deconfined quarks and gluons propagate through the QGP medium and lose energy. This results in a modification to the yields of several particle species relative to their yields in $p+p$ collisions, with the notable exception of the direct photon, which escapes the plasma unmodified. One area of research is quantifying how energy is dissipated through the QGP via either gluon radiation or by secondary partonic collisions within the QGP. These proceedings detail new results on energy loss studies from PHENIX using two techniques: measurements of the single particle $R_{AA}$ and two-particle correlations. 

\section{Single Particle $R_{AA}$}
The $R_{AA}$, or nuclear modification factor, of a given particle species quantifies the extent to which the yield of a given particle species in $A+A$ collisions is modified relative to the yield in $p+p$ scaled by the number of binary collisions. The  definition of $R_{AA}$ is given in Eqn. \eqref{eq:1}.
 %$R_{AA}$ is the ratio between the yield of a given particle species in $A+A$ to the yield of the same species in $p+p$ scaled to the same $N_{Coll}$, as shown in equation \eqref{eq:1}.

\begin{equation}
    %R_{AA} = \frac{\sigma^{NN}}{\langle N_{Coll} \rangle}\frac{d^2 N^{AA}/ dp_T d_{\eta}}{d^2 \sigma_{pp}/dp_T d_{\eta}} \label{eq:1}
    R_{AA} = \frac{1}{\langle T_{AA} \rangle}\frac{d^2 N^{AA}/ dp_T d{\eta}}{d^2 \sigma_{pp}/dp_T d{\eta}} \label{eq:1}
\end{equation}

%A new PHENIX result showing the comparison between the proton $R_{AB}$ and $R_{AA}$ from $Cu+Au$ and $Au+Au$ at $\sqrt{s_{NN}}=200$~GeV is shown in Fig. \ref{Cu_Au_p_RAB}. The $Au+Au$ $R_{AA}$ result was previously shown in \cite{Au_Au_RAA_for_Cu_Au}. In the most central collisions at moderate $p_T$, there appears an enhancement in the yield of protons that is less prominent for more peripheral collisions. For similar system size (or $N_{Part}$, which is the number of participating nucleons in a collision), the magnitude of the enhancement in the proton yield $Au+Au$ is nearly the same as that in $Cu+Au$. The fact that the proton $Cu+Au$ and $Au+Au$ match at similar $N_{Part}$ means that the observed effect, quark recombination, is dependent on the centrality of the collision, rather than the collision species itself. 

%%\begin{figure}[h]
%%    \centering
%%    \includegraphics[scale=0.4]{QM19-NPA-template/Figures/protons_CuAuAuAu.png}
%%    \caption{Proton $R_{AB}$ in $Cu+Au$ (green) and $Au+Au$ (purple) collisions. The leftmost panel represents central collisions (high $N_{part}$), and the subsequent panels represent more and more peripheral (lower $N_{Part}$) collisions.}
%%    \label{Cu_Au_p_RAB}
%%\end{figure}

\begin{figure}[!h]
   \begin{floatrow}
\ffigbox{\includegraphics[width=0.75\columnwidth]{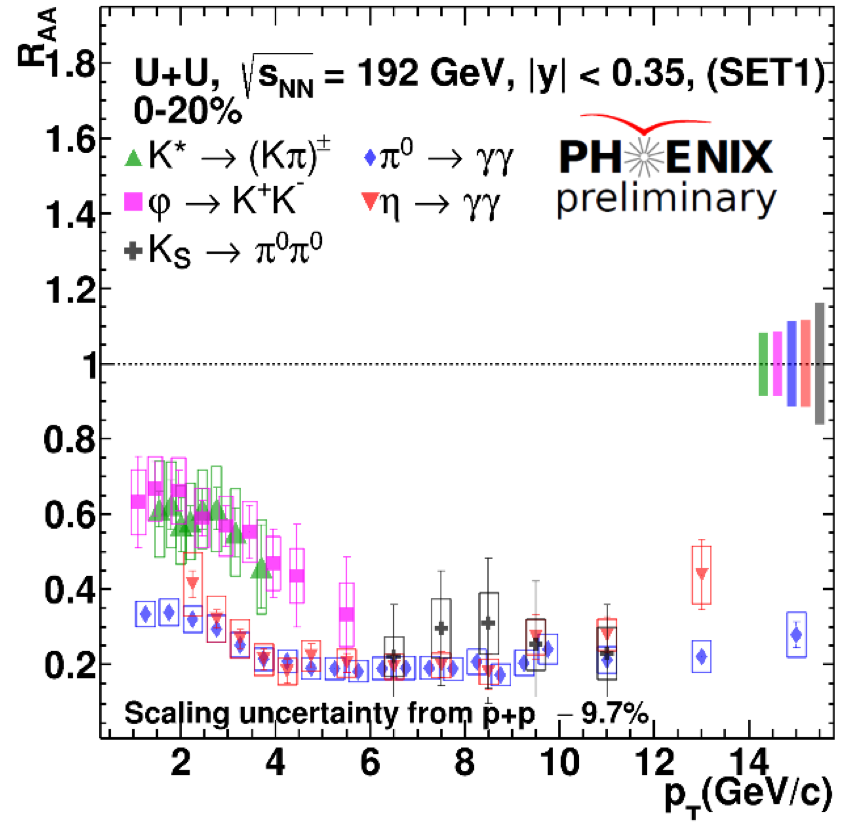}}%
        {\caption{$R_{AA}$ for various meson species as a function of $p_T$ in $U+U$ at $192$~GeV collisions. One can see features of possible strangeness enhancement at low $p_T$ in the 0--20$\%$ centrality bin which disappears at higher $p_T$}\label{U_U_Cu_Au_Mesons}}
\hfill
\ffigbox{\includegraphics[width=\columnwidth]{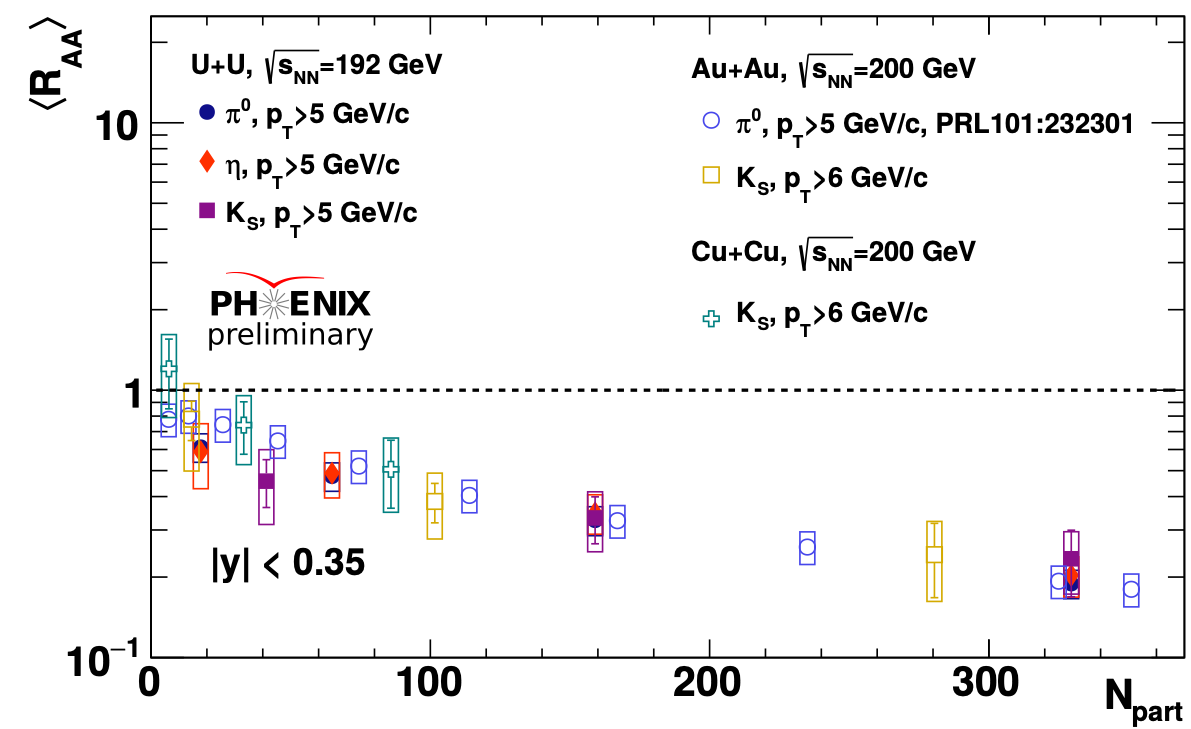}}%
        {\caption{Integrated $R_{AA}$'s for different meson species and collision systems for $p_T > 5$~GeV/c}\label{Int_RAA}}
   \end{floatrow}
\end{figure}

PHENIX has measured the $R_{AA}$ for several meson species in $Cu+Au$ at $\sqrt{s_{NN}}=200$~GeV and $U+U$ $\sqrt{s_{NN}}=192$~GeV, with the $U+U$ results shown in Fig. \ref{U_U_Cu_Au_Mesons}. The $R_{AA}$ is plotted on the $y$-axis as a function of the transverse $p_T$ on the $x$-axis. In both collision systems in the most central collisions, there is a separation between the $\phi$ and $K^{*}$ $R_{AA}$ and the $\pi^0$ and the $\eta$ ~$R_{AA}$ at low $p_T$. All species remain suppressed relative to the $p+p$ baseline across all $p_T$, however. This separation may be attributable to strangeness enhancement which reflects the quark content of the species shown. The $\phi$, for instance, is entirely composed of strange quarks $(s\bar{s})$, whereas the $\pi^0$ ($((u\bar{u})+(d\bar{d}))/\sqrt{2}$) has no contribution to its makeup from strange quarks and, thus, would not see this effect. The dominant source of systematic uncertainty (represented by the empty boxes around the data points) is the uncertainty in the raw yield extraction and in the reconstruction efficiency.

\vspace{-3mm}
The $p_T$ integrated $\langle R_{AA} \rangle$ across $Au+Au$, $Cu+Au$, and $U+U$ as a function of the system size $N_{part}$ and for various mesons is shown in Fig. \ref{Int_RAA}, with the $\langle R_{AA} \rangle$ on the $y$-axis and the transverse $p_T$ on the $x$-axis. The different data points correspond to different particle species. When only looking at the high $p_T$ region ($p_T > 5$~GeV/c), effects due to strangeness enhancement disappear, and the integrated $\langle R_{AA} \rangle$ for all meson species across the different collision species follow the same trend. This shows that at high $p_T$, energy loss is the dominant QGP-related effect. 

\section{Two-Particle Correlations}
Measurement of two-particle correlations allows one to study the modification of both the particle yield and shape of recoil jets in $A+A$ relative to a $p+p$ baseline. In this method, we correlate all the charged hadrons in an event to a high $p_T$ trigger particle (in this case, the $\pi^0$) and measure the angular separation in the azimuthal direction, $\Delta \phi$. This results in a distribution with two prominent peaks at $\Delta \phi =0$ and at $\Delta \phi = \pi$, which are called the near and away-side peaks, respectively. From here, subtraction of correlations due to flow (in this analysis flow harmonics up to $n=4$ are subtracted) yields the jet function, defined in Eqn. \ref{jetfunctioneqn}. 

\begin{equation}
    \frac{1}{N^t}\frac{dN^{Pair}}{d\Delta \phi}= \frac{1}{N^t} \frac{N^{Pair}}{\epsilon\int\Delta\phi} \left\{ \frac{dN^{Pair}_{\mathit{Same}}/d\Delta\phi}{dN^{Pair}_{Mix}/d\Delta\phi} - \xi(1+\sum_{n=2}^4 2\langle v_n^t\rangle \langle v_n^a \rangle \cos(n\Delta\phi))\right\} \label{jetfunctioneqn}
\end{equation}

\noindent Here, $N^t$ and $N^{Pair}$ refer to the number of trigger $\pi^0$'s and correlated pairs, respectively. ``Same" denotes a $\pi^0$-hadron pair that come from the same event, whereas pairs denoted ``Mix" come from two separate events and are used to correct for detector effects via event mixing. $\epsilon$ is the charged hadron efficiency, and $\xi$ represents the magnitude of the correlated background due to flow, calculated using the absolute background subtraction method \cite{AbsNorm}. Lastly, $v_n^t$ and $v_n^a$ represent the flow harmonic values for both the trigger and the associated hadron, respectively. Both are taken from previous PHENIX analyses \cite{pi0_vn_ref,had_vn_ref}.

\begin{figure}[!h]
   \begin{floatrow}
\ffigbox{\includegraphics[width=\columnwidth]{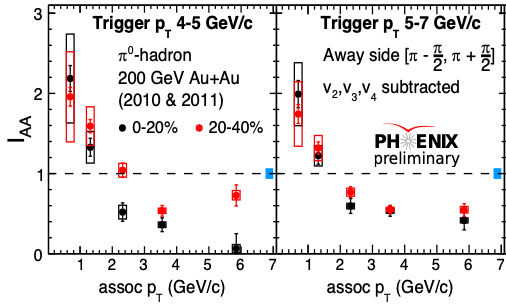}}%
        {\vspace{-3mm}
        \caption{$I_{AA}$ as a function of the associate particle $p_T$ in two trigger $p_T$ bins, and in 0--20$\%$ (black) and 20--40$\%$ (red) centrality classes.}\label{IAA_pt}}
\ffigbox{\includegraphics[width=\columnwidth]{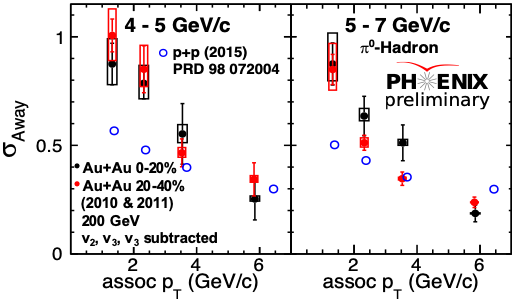}}%
        {\vspace{-3mm}
        \caption{Extracted away-side Gaussian widths plotted as a function of the associate partilce $p_T$ in 0--20$\%$ (black) and 20--40$\%$ (red) centrality classes. The blue points are the $p+p$ baseline.}\label{jetwidth}}
   \end{floatrow}
\end{figure}

\begin{figure}[H]
    \begin{center}
    \includegraphics[width=\textwidth]{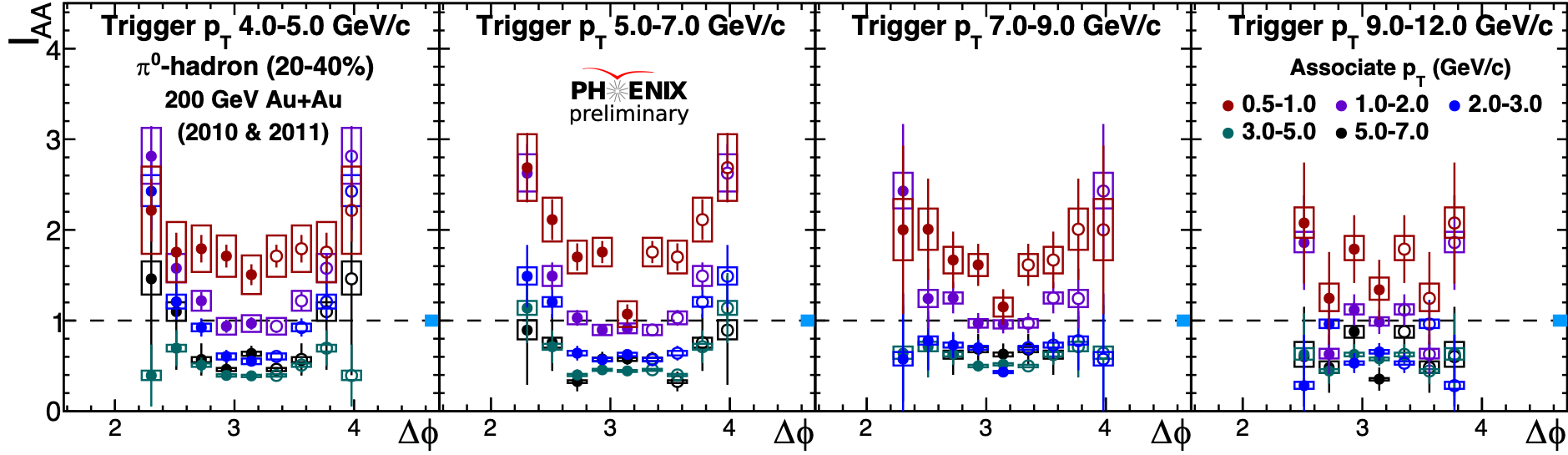}
    \vspace{-10mm}
    \caption{$I_{AA}$ as a function of the separation angle $\Delta \phi$ in four trigger $p_T$ bins. Each color represents a different associate particle $p_T$ range. The filled points are calculated directly by dividing a given $Au+Au$ jet function by the same bin in the $p+p$ baseline. These points are then mirrored across $\Delta \phi = \pi$, which is represented by the open points.}
    \label{IAA_DPhi}
    \end{center}
\end{figure}

From the jet function, we extract the yield of the away-side jet peak by integrating the region \mbox{$|\Delta \phi - \pi| < \pi/2$}. We then take the integrated yields in $A+A$ and $p+p$, $Y_{AA}$ and $Y_{pp}$, respectively, and calculate the \mbox{$I_{AA} = Y^{AA}/Y^{pp}$}. Fig. \ref{IAA_pt} shows the $I_{AA}$ on the $y$-axis plotted as a function of the associate particle $p_T$ on the $x$-axis in two different trigger $p_T$ ranges. The empty boxes around the data points in Fig. \ref{IAA_pt}, \ref{jetwidth}, and \ref{IAA_DPhi} represent the systematic uncertainties and stem from the $\pi^0$ combinatorial background, the estimation of the background level $\xi$, and from the flow harmonic measurements themselves from \cite{pi0_vn_ref,had_vn_ref}. Additionally, the blue box shown at unity in Fig. \ref{IAA_pt} and \ref{IAA_DPhi} represents a global scale uncertainty primarily coming from the charged hadron tracking efficiency correction, $\epsilon$.  $I_{AA}$ is an important observable in two-particle correlations because it allows us to directly study modifications to the fragmentation function, $D(z)$, due to the fact that the integrated yields are approximately equivalent to the  fragmentation functions; that is to say: $I_{AA}=Y^{AA}/Y^{pp}\approx D^{AA}(z)/D^{pp}(z)$.

%
%\begin{figure}[H]
%    \begin{center}
%    \includegraphics[width=\textwidth]{QM19-NPA-template/Figures/IAA_pt_QM2019.png}
%    \caption{$I_{AA}$ as a function of the associate particle $p_T$ in four trigger $p_T$ bins, and in 0--20$\%$ (black) and 20--40$\%$ %(red) centrality classes.}
%    \label{IAA_pt}
%    \end{center}
%\end{figure}

The $I_{AA}$ measurement shown in Fig. \ref{IAA_pt} shows an enhancement in the yield of low $p_T$ associate particles, and a suppression in the yield of high $p_T$ associate particles. This trend is consistent across the two trigger $p_T$ bins shown and in the 0--20$\%$ and 20--40$\%$ centrality bins. Additionally, the width of the away-side jet peak was measured by fitting it with a Gaussian and extracting the width, $\sigma$, as previously shown in \cite{Connors_HP_2019}. The result of this measurement is shown in Fig. \ref{jetwidth}, with the jet width, $\sigma$, on the $y$-axis and the associate particle $p_T$ on the $x$-axis. The away-side widths in $Au+Au$ collisions are broader than those in $p+p$ at low associate particle $p_T$. At high $p_T$, the $Au+Au$ jet widths converge to the $p+p$ baseline in both centrality bins. 

The broadening of the away-side at low $p_T$ seen in \cite{Connors_HP_2019} coupled with the enhancement in the away-side yield at low $p_T$ in Fig. \ref{IAA_pt} show us that the mechanism behind energy loss within the QGP is consistent with gluon bremsstrahlung being radiated at wide angles relative to the recoil parton trajectory. We extend this analysis further to track the lost energy in $\Delta\phi$ space. To do this, we plot the $I_{AA}$ in a given trigger $p_T$ bin as a function of the separation angle $\Delta \phi$ over the range $|\Delta \phi - \pi| \lesssim \pi/3$, encompassing the away-side jet peak. The result is shown in Fig. \ref{IAA_DPhi}. Each of the different sets of color points represents a different associate particle $p_T$ range, with red being the lowest ($0.5$--$1$~GeV/c) and black being the highest ($5$--$7$~GeV/c). The $I_{AA}$ for the highest $p_T$ associate particles, (i.e. the $5$--$7$~GeV/c and $3$--$5$~GeV/c bins), is suppressed by nearly the same amount across the $\Delta\phi$ range. As the $p_T$ of the associate particle decreases, however, the $I_{AA}$ begins to take on a $\Delta \phi$ dependence, showing enhancement at wide angles relative to the away-side jet peak ($\Delta \phi = \pi$), as is seen in the \mbox{$0.5$--$1$~GeV/c} and $1$--$2$~GeV/c bins. Near $\Delta \phi \approx \pi$, however, the $I_{AA}$ for the $1$--$2$~GeV/c bin appears to be consistent with one, whereas the $0.5$--$1$~Gev/c shows enhancement across almost all $\Delta \phi$ bins on the away-side. Lastly, there appear to be no major changes in the measurement when looking at different trigger $p_T$ bins. 

\section{Conclusions}

Single particle $R_{AA}$ measurements have been performed by PHENIX in $A+A(B)$ collisions at $\sqrt{s_{NN}}=200$~GeV and $193$~GeV. At high $p_T$, the integrated $R_{AA}$'s for several meson species were found to follow the same trend as a function of $N_{Part}$, showing that the modification in their yield is dependent only on $N_{Part}$ and not the collision species nor meson species. This means that modifications to single particle yields at high $p_T$ are dominantly due to energy loss.

Measurements of $\pi^0$-triggered two-particle correlations from PHENIX in $Au+Au$ collisions at $\sqrt{s_{NN}}=200$~GeV from RHIC Year-10 and 11 datasets show modification to the away-side jet peak's width and yield relative to the $p+p$ baseline. Extraction of the $I_{AA}$ and jet width $\sigma$ show that, at low associate particle $p_T$, recoil jets in $Au+Au$ are broader and have a larger particle yield than in $p+p$. Meanwhile, at high associate particle $p_T$, these jets are as collimated as their $p+p$ counterparts, and their particle yield is suppressed relative to the $p+p$ baseline. Lastly, a new measurement, the $I_{AA}$ as a function of $\Delta \phi$, shows that high $p_T$ associate particles are suppressed at the same level in $\Delta \phi$ space, whereas the enhancement of low $p_T$ associate particles has a $\Delta \phi$ dependence and is most prominent at large angles relative to the away-side jet peak. The two-particle correlation studies presented in this analysis will be expanded upon by adding the statistics of two of PHENIX's largest $Au+Au$ at $\sqrt{s_{NN}}=200$~GeV datasets, Run 14 and Run 16, as well as including direct photon triggered correlations. 

%% The Appendices part is started with the command \appendix;
%% appendix sections are then done as normal sections
%% \appendix

%% \section{}
%% \label{}

%% References
%%
%% Following citation commands can be used in the body text:
%% Usage of \cite is as follows:
%%   \cite{key}         ==>>  [#]
%%   \cite[chap. 2]{key} ==>> [#, chap. 2]
%%

%% References with BibTeX database:
\section*{Acknowledgements}
This work was supported by the National Science Foundation [Grant Number~1714801 \& 1848162]
\bibliographystyle{elsarticle-num}
\bibliography{main.bib}

%\end{thebibliography}

%% Authors are advised to use a BibTeX database file for their reference list.
%% The provided style file elsarticle-num.bst formats references in the required Procedia style

%% For references without a BibTeX database:

% \begin{thebibliography}{00}

%% \bibitem must have the following form:
%%   \bibitem{key}...
%%

% \bibitem{}

% \end{thebibliography}

\end{document}